\documentclass[%
 prl,twocolumn,
 superscriptaddress,
 showpacs,
 longbibliography,
 aps,
 showkeys,
]{revtex4-1}

\usepackage[pdftex]{hyperref,color,graphicx}
\usepackage{amsfonts,amssymb,amsmath}
\usepackage[separate-uncertainty=false]{siunitx}
\usepackage[english]{babel}
\usepackage[utf8]{inputenc}
\usepackage[T1]{fontenc}

\usepackage{graphicx}
\usepackage[space]{grffile}
\usepackage{latexsym}
\usepackage{textcomp}
\usepackage{longtable}
\usepackage{multirow,booktabs}
\usepackage{url}
\hypersetup{colorlinks=false,pdfborder={0 0 0}}
\usepackage[english]{babel}
\usepackage{dcolumn}
\usepackage{bm}
\usepackage{units}
\usepackage{upgreek}
\usepackage{color}
\usepackage{cancel}
\usepackage{hyperref}
\usepackage{braket}

\begin{document}

\title{Quantum Walk in Momentum Space with a Bose-Einstein Condensate}

\author{Siamak Dadras}
\affiliation{Department of Physics, Oklahoma State University, Stillwater, Oklahoma 74078-3072, USA}

\author{Alexander Gresch}
\affiliation{ITP, Heidelberg University, Philosophenweg 12, 69120 Heidelberg, Germany}
\author{Caspar Groiseau}
\affiliation{ITP, Heidelberg University, Philosophenweg 12, 69120 Heidelberg, Germany}

\author{Sandro Wimberger}
\affiliation{Dipartimento di Scienze Matematiche, Fisiche e Informatiche, Universit\`{a} di Parma, Parco Area delle Scienze 7/A, 43124 Parma, Italy}
\affiliation{INFN, Sezione di Milano Bicocca, Gruppo Collegato di Parma, Parma, Italy}
\affiliation{ITP, Heidelberg University, Philosophenweg 12, 69120 Heidelberg, Germany}

\author{Gil S. Summy}
\email{gil.summy@okstate.edu}
\affiliation{Department of Physics, Oklahoma State University, Stillwater, Oklahoma 74078-3072, USA}

\begin{abstract}
We present a discrete-time, one-dimensional quantum walk based on the entanglement between the momentum of ultracold rubidium atoms (the walk space) and two internal atomic states (the ``coin'' degree of freedom). Our scheme is highly flexible and can provide a platform for a wide range of applications such as quantum search algorithms, the observation of topological phases, and the realization of walks with higher dimensionality. Along with the investigation of the quantum-to-classical transition, we demonstrate the distinctive features of a quantum walk and contrast them to those of its classical counterpart. Also, by manipulating either the walk or coin operator, we show how the walk dynamics can be steered or even reversed.\end{abstract}




\maketitle

Quantum randomness is intrinsically different from classical stochasticity since it is affected by interference and entanglement. Entanglement is responsible for nonlocal correlations \cite{EPR1935} and is the resource of quantum computing \cite{NJ2000}. While the basic procedure for producing a quantum walk (QW) can be outwardly similar to its classical counterpart, the dynamics of a QW are completely different and can lead to applications unavailable classically. Notably, QWs are intrinsically connected to quantum search algorithms (see, e.g., \cite{Book_QS}) and to quantum algorithms in general \cite{DT-QC, Childs2013}.

Several different experimental QW schemes have been implemented. Walks have been carried out with atoms \cite{PNAS2012, Topo2016, Preiss2015, meschede2009}, ions \cite{PhysRevLett.103.090504,PhysRevLett.104.100503}, or photons \cite{Topo2,PhysRevLett.104.153602,Silberberg2010, PhysRevLett.100.170506, PhysRevLett.114.203602}. As might be expected, the variety of different possible walker species leads to the walks themselves taking on an assortment of different forms. For example, walks with photons are most conveniently done in the time domain \cite{Silberberg2010, PhysRevLett.100.170506} or more recently in angular momentum \cite{Topo2}, while for atoms and ions the walks are usually performed in spatial \cite{meschede2009,Preiss2015} or phase \cite{PhysRevLett.103.090504,PhysRevLett.104.100503} degrees of freedom (d.o.f.). However, up to now, no experimental realization of a QW has been reported in momentum space, which we will argue has several important benefits. Our QW offers distinct advantages arising from the robustness of its dynamics in momentum space and extendability to higher dimensions  \cite{Science2012, PhysRevLett.112.143604, sadgrove2012} and many-body regimes \cite{Preiss2015,Spin1,AW2017}.

The discrete-time QW we describe here consists of two d.o.f.: the walker’s space (momentum space in our case) and a “coin” which selects the path of the system through the walker’s space. The key concept that differentiates a classical walk from a QW is that in the latter case there exists a strong entanglement between the d.o.f. This entanglement leads to distinct behavior which is the result of the interference between the multitude of paths that a walker may take simultaneously in the walk space. For example, this produces one of the characteristic signatures of a QW, the appearance of two peaks in the walk distribution that propagate ballistically away from the origin as the walk proceeds.

In this Letter we demonstrate the principal features of QWs resulting from interference and contrast them to the behavior of a classical walk. By manipulating either the walk or coin operator we show how the walk can be biased or reversed. Future applications can build on the implicit spin-momentum coupling of our walk that is also a necessary ingredient for studying topological effects \cite{Gong2012,Tian2014,Gong2018}.

The implementation of our QW is carried out with a Bose-Einstein condensate (BEC) of $^{87}$Rb atoms in a pulsed optical lattice. One of the major benefits of a momentum-based QW is that it provides straightforward access to both internal and external d.o.f. of the walker. In our system these d.o.f. are two atomic hyperfine states and the center-of-mass momentum of the atoms.

Each step of a discrete-time walk ${\hat{\bf{U}}}_{\rm{step}}={\hat{\bf{T}}}{\hat{\bf{M}}}$ consists of a coin operator ${\hat{\bf{M}}}$ which produces a superposition of two internal states, followed by a unitary shift operator ${\hat{\bf{T}}}$, whose direction is determined by the internal state. We realize the coin operator,
\begin{equation}
{\hat{\bf{M}}}({\alpha},{\chi})=
\begin{bmatrix}
& \cos{({\alpha}/2)}& {e^{-i\chi}}{\sin({\alpha}/2)}\\
& {-e^{i\chi}}{\sin({\alpha}/2)} & \cos{({\alpha}/2)}\\
\end{bmatrix},
 \label{eq:coin} 
\end{equation}
using resonant microwave (MW) radiation that addresses the internal d.o.f., the ground hyperfine level components $F=1$, $m_{F}=0$ and $F=2$, $m_{F}=0$. Henceforth these states are denoted by $|1\rangle$ and $|2\rangle$, respectively. In a regular coin toss operation, a ${\pi}/2$ MW pulse ${\hat{\bf{M}}}(\pi/2,-\pi/2)$ produces an equal superposition of internal states $1/{\sqrt{2}}(|1\rangle +i|2\rangle)$ at each step of the walk. To make the direction of the walk contingent upon the internal d.o.f., we apply the unitary shift operator, 
\begin{equation}
\hat{\bf{T}} = \exp{(iq{\hat{\theta}})} |1\rangle \langle 1| + \exp{(-iq{\hat{\theta}})} |2\rangle \langle 2|,
 \label{eq:Shift} 
\end{equation}
which changes the momentum by $\pm{q}$ depending on whether the atom resides in the internal state $|1\rangle$ or $|2\rangle$. This produces a strong entanglement between internal and external d.o.f. at each step of the walk. In Eq.~\eqref{eq:Shift}, $q$ is an integer in units of two-photon recoils ${\hbar}G$ ($G=2\pi/\lambda_G$ with $\lambda_G$ being the spatial period of the pulsed optical lattice implementing the momentum walk) and ${\hat{\theta}}=\hat{x}\mod(2\pi)$, where $\hat{x}$ is the dimensionless position operator. In the usual walk, $q=1$, corresponding to nearest neighbor coupling in momentum space. 

As proposed in \cite{Gil2016a}, our shift operator is a quantum ratchet derived from the atom-optics kicked rotor (AOKR) \cite{RaizenAdv,SW2011}. AOKR experiments work with ultracold atoms subject to a series of short pulses of a 1D off-resonant optical lattice (standing wave). Using dimensionless variables, the dynamics of the AOKR are described by the single-particle Hamiltonian \cite{RaizenAdv}, 
\begin{equation}
\label{eq:3}
{\hat{H}}({\hat{x}},{\hat{p}}_{x},t)={{\hat{p}}_{x}}^2/2+k\cos({\hat{x}})\sum_{j{\in}{\mathbb{Z}}}{\delta}(t-j\tau),
\end{equation}
where $j$ counts the number of pulses, $t$ and $\tau$ are the dimensionless time and pulse period, and ${\hat{p}}_{x}$ is the rescaled momentum operator.  The kick strength is $k={\Omega}^2{\tau_p}/\Delta$, in which ${\tau_p}({\ll}\tau)$ is the pulse length, $\Omega$ is the Rabi frequency, and $\Delta$ is the detuning of the laser light from the atomic transition. The evolution of the system during the pulse and subsequent free evolution is given by the Floquet operator
${\hat{\mathcal{U}}}_{{\hat{p}_{x}},k}={\hat{\mathcal{U}}}_f{\hat{\mathcal{U}}}_k
=e^{-i{\tau}{{\hat{p}}_{x}}^2/2}e^{-ik{\cos}({\hat{\theta}})}$. Here, ${\hat{\mathcal{U}}}_f$ signifies the free evolution between two pulses, and ${\hat{\mathcal{U}}}_k=\sum_{m}(-i)^{m}{\exp}(-im{\hat\theta)}J_{m}(k)$. The $J$'s are Bessel functions of the first kind and represent behavior in which each pulse leads to a symmetric diffraction of the wave function in the space spanned by the momentum eigenstates $\{|n\rangle\}, n\in \mathbb{Z}$. To realize the simplest ratchets, the dynamics of the AOKR should meet the quantum resonance conditions \cite{SW2011}. This can be understood as the Talbot effect (in the time domain) for matter waves diffracted from a phase grating induced by a pulsed optical standing wave \cite{SW2011}. The Talbot condition is realized by choosing $\tau=4\pi$ for our dimensionless pulse period in Eq.~\eqref{eq:3}. Experimentally a Talbot time must be long enough to allow for the delivery of the coin-toss MW radiation between the ratchet pulses. 

A ratchet can be created by breaking the spatial-temporal symmetry of the usual AOKR \cite{SW2011,Hoogerland2013} through the choice of a particular initial external state. Experimentally, the simplest choice is  $1/\sqrt{2}(|n=0\rangle+e^{i\phi}|n=1\rangle)$ realized with a long pulse of the off-resonant standing wave (Bragg pulse) on the original BEC ($|n=0\rangle$) \cite{Gil2016b, Ann2017}. Subsequent application of the AOKR to this state results in the average momentum changing by an amount ${\Delta}\langle{\hat{p}}\rangle=-k\sin{(\phi)}/2$ at each AOKR pulse \cite{SW2011}. By choosing $\phi=\pi/2$ and $|k|{\sim}2$, we either decrease or increase (depending on the sign of $k$) the average momentum at each step of the AOKR (or now ratchet) by one unit. Recall that $k\propto{1/\Delta}$, so that with the light detuned with positive $\Delta$ for $|1\rangle$ and negative $\Delta$ for $|2\rangle$, it is possible for the internal states to undergo simultaneous ratchets in opposite directions.

In order to implement a standard QW with $j$ steps, we apply the sequence $(\hat{\bf{U}}_{\rm{step}})^j=[{\hat{\bf{T}}}{\hat{\bf{M}}}(\pi/2,-\pi/2)]^{j-1}[{\hat{\bf{T}}}{\hat{\bf{M}}}(\pi/2,\pi)]$ to the initial state $|\psi_0\rangle=|1\rangle\otimes1/\sqrt{2}(|n=0\rangle+ i|n=1\rangle)$ prepared by a Bragg pulse. The first step includes the ${\hat{\bf{M}}}(\pi/2,\pi)$  coin toss, the well known Hadamard gate, that we use to prepare the internal states to perform a symmetric walk. Our observable is the momentum distribution, represented by the atomic population of momentum states, $P(n)$. When the state of the system in momentum space after $j$ walk steps is $|{\psi}(j){\rangle}=\sum_{n}c_{n}(j)|n\rangle$, we measure the momentum distribution
$P(n,j)=P_{|1\rangle}(n,j)+P_{|2\rangle}(n,j)={|c_{n,1}(j)|}^2+{|c_{n,2}(j)|}^2$,
containing the population distributions of both internal states. 

\begin{figure}
	\begin{center}
		\centering
		\includegraphics[scale=0.0553]{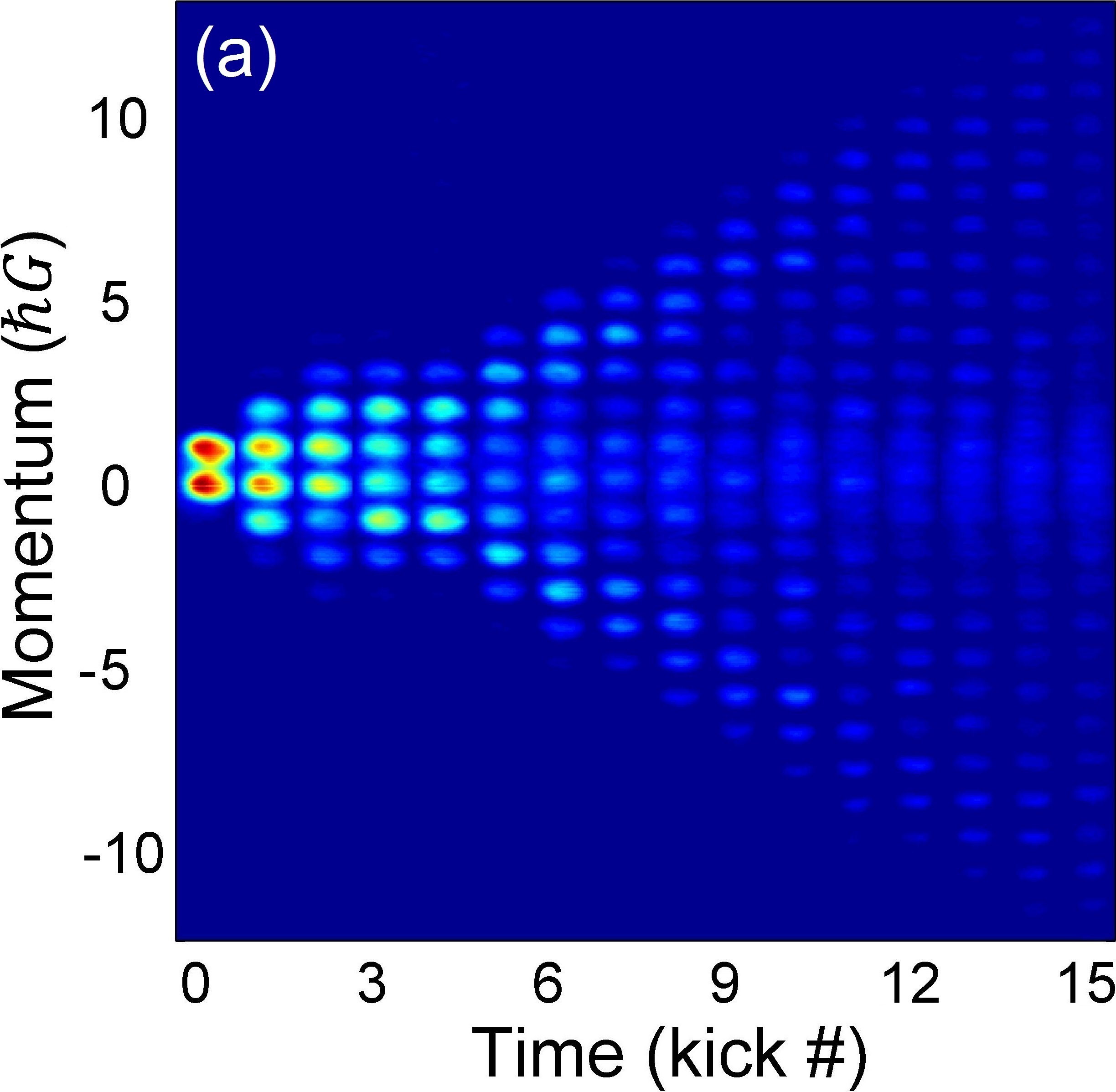}
		\hspace{0.05cm}
		\includegraphics[scale= 0.0545]{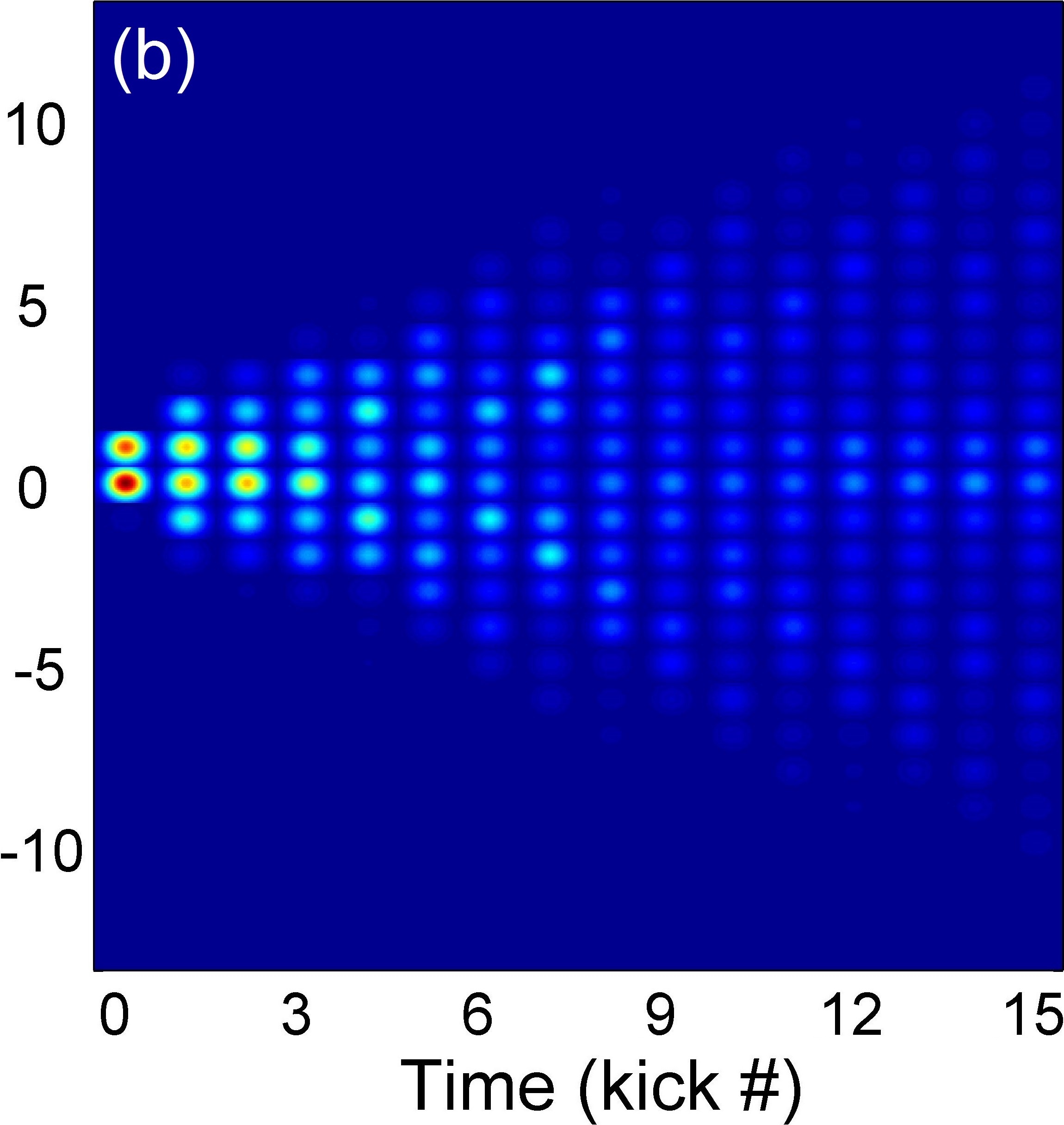}\\
		\vspace{0.2cm}
		\includegraphics[scale=0.052]{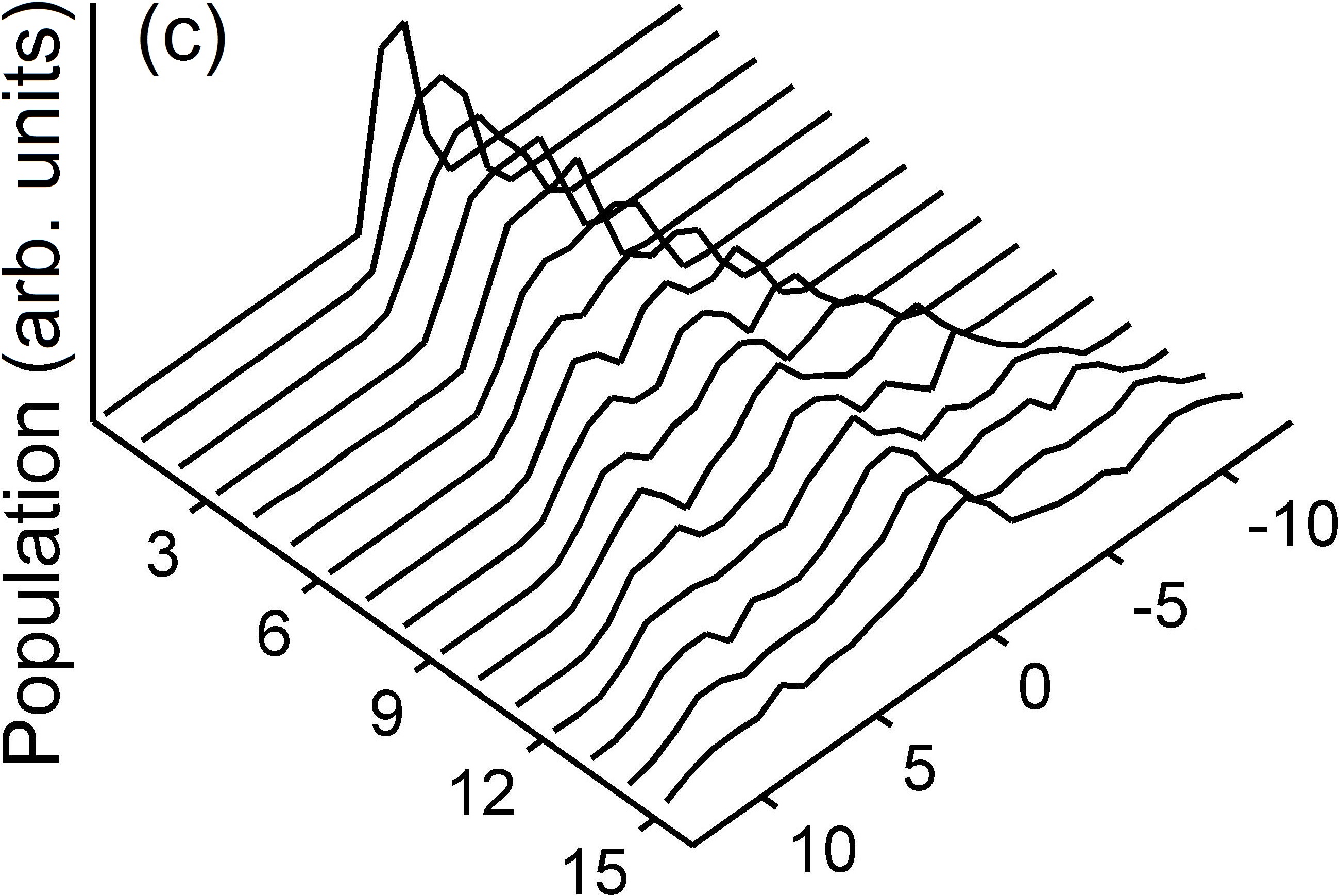}
		\hspace{0.1cm}
		\includegraphics[scale= 0.052]{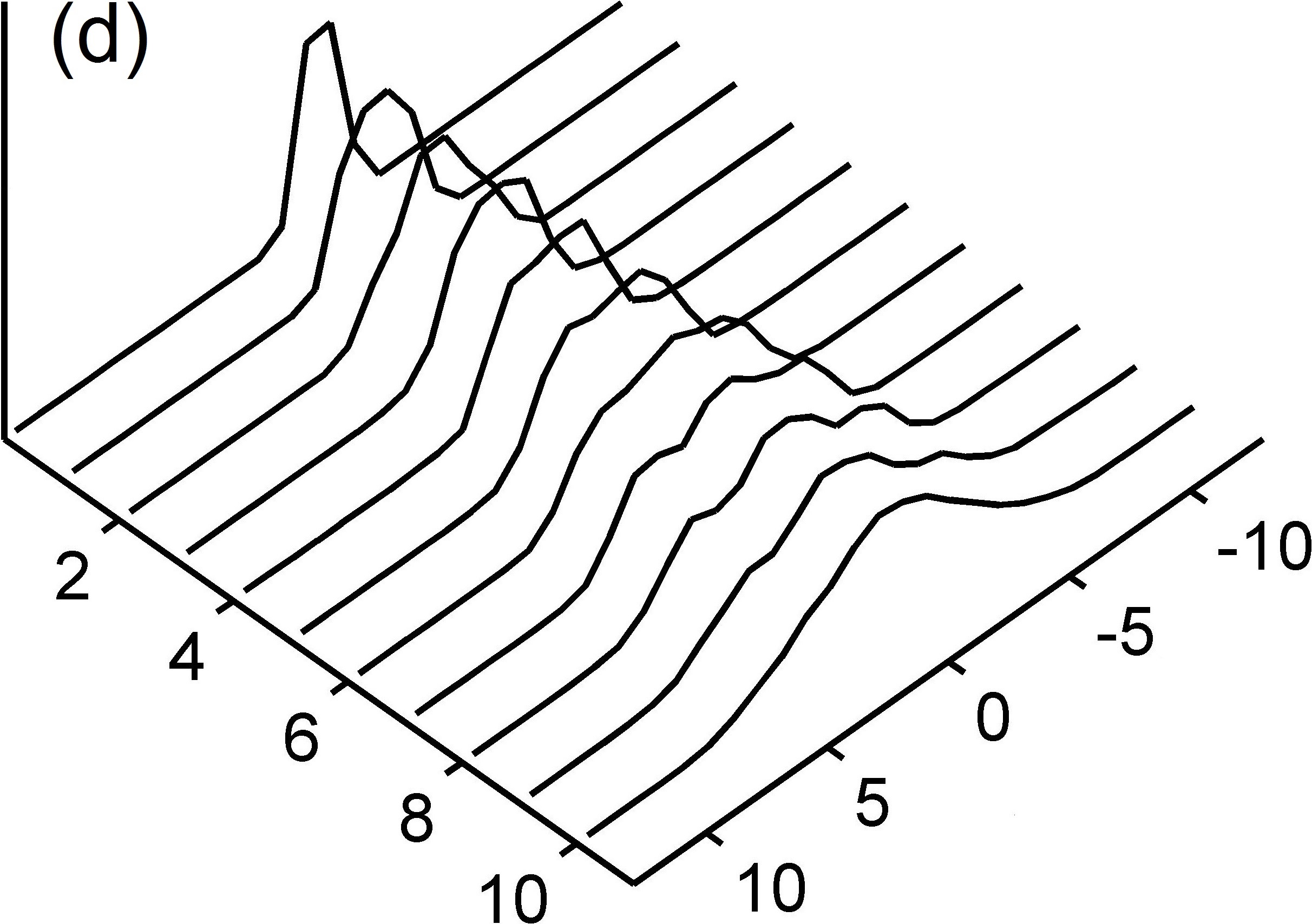}\\
		\vspace{0.2cm}
		\includegraphics[scale=0.052]{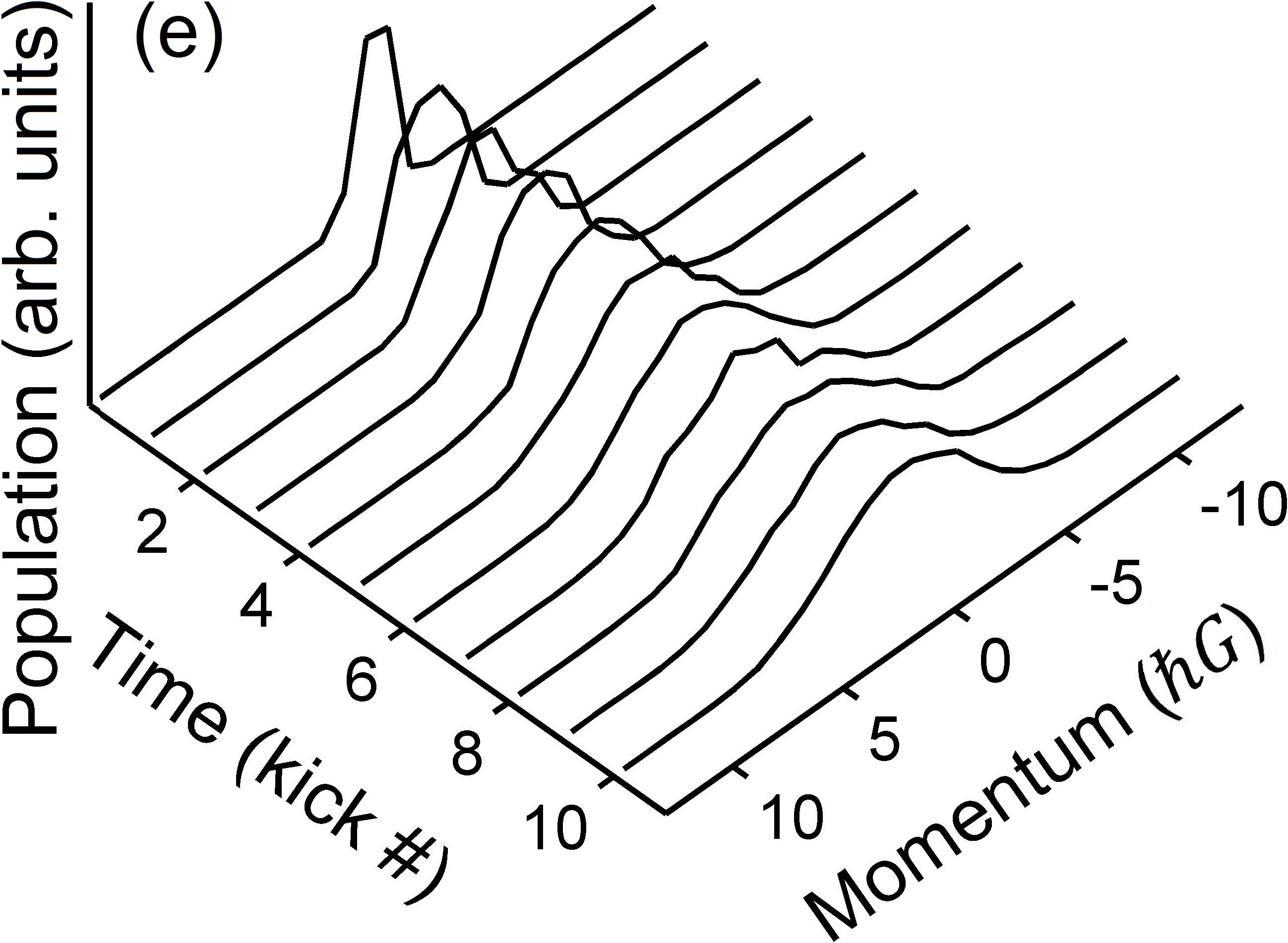}
		\hspace{0.1cm}
		\includegraphics[scale=0.052]{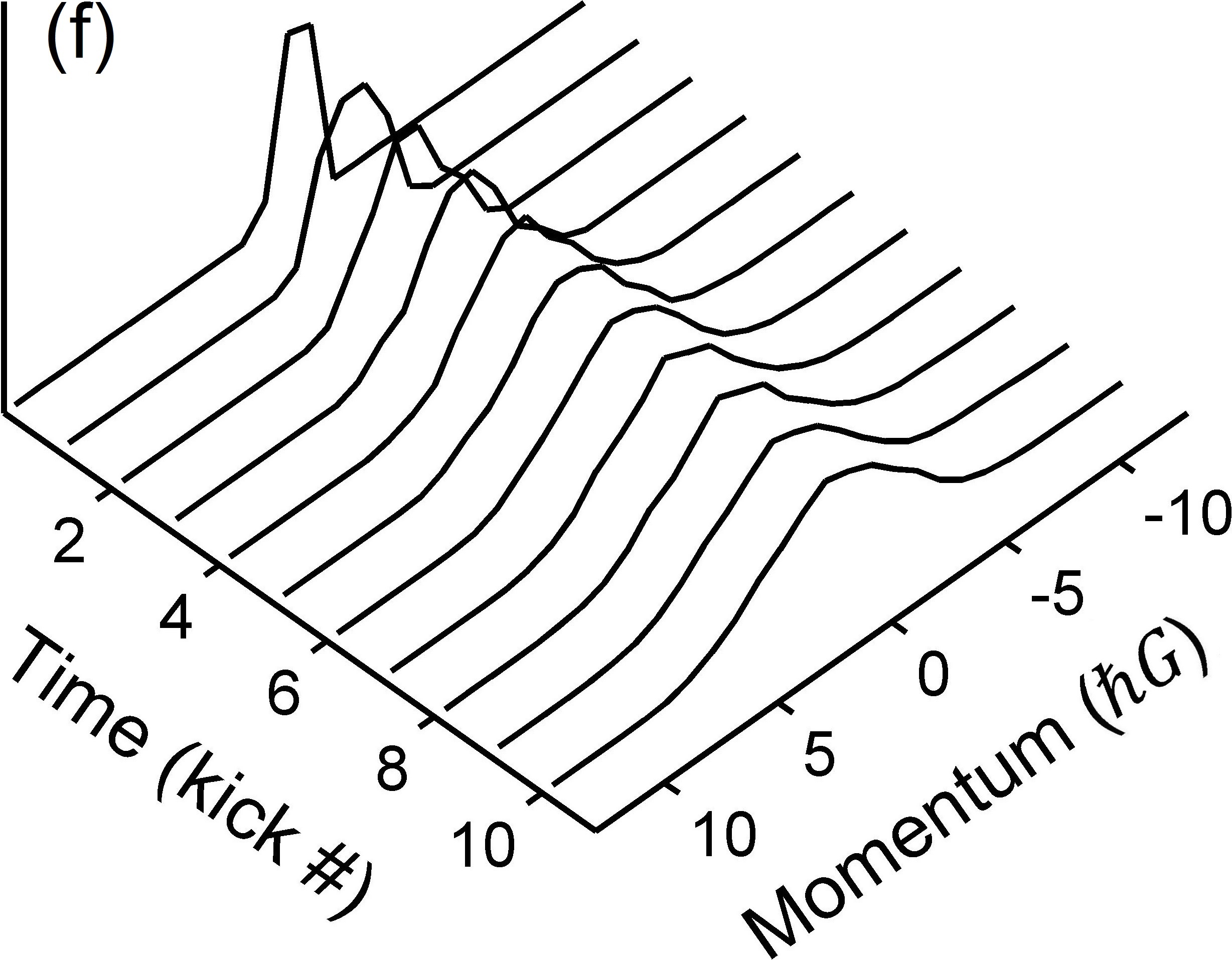}\\
		\vspace{0.2cm}
		\includegraphics[width=0.7\linewidth]{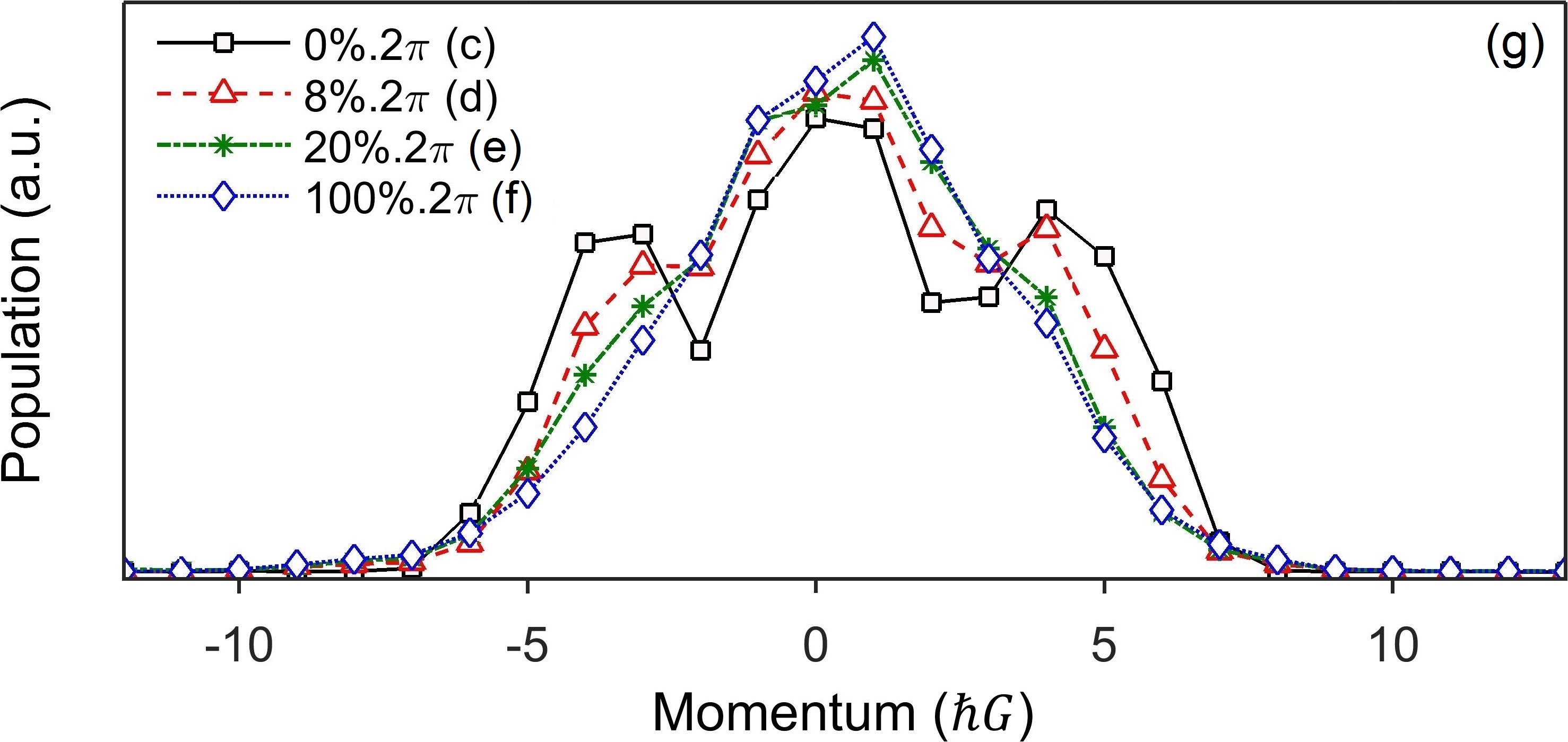}\\
\caption{Experimental (a) and simulated (b) momentum distributions of our standard QW realized with $|k|=1.45$. Each time (kick $\#$) represents one step of the walk, i.e., one realization of the experiment (or simulation). The amorphous population signal about the center of the momentum distribution in (a) is a residual atomic thermal cloud that, unlike the BEC, does not respond to the ratchet. The middle panels represent the quantum-to-classical transition: the standard QW (c) was conducted with a fixed coin toss phase. Signatures of a classical walk emerge at $8\%$ phase noise (d). The walk is dominantly classical when randomizing the phase by $20\%$ (e) and becomes fully classical when the phase is allowed to vary randomly within $2\pi$ (f). Panel (g) shows the evolution of momentum distribution pattern at the eighth step of the walk for the corresponding noise levels.} \label{fig1}
	\end{center}
\end{figure}

Figures~\ref{fig1}(a) and \ref{fig1}(b) show experimental and simulated results for the momentum distribution of our QW realized with the shift operator strength $|k|=1.45$. Among various realizations of QWs, this ratchet strength best matches the standard QW by coupling neighboring momentum states [with $q=1$ in Eq.~\eqref{eq:Shift}]. Thus, we employed this value of $k$ in the subsequent experiments. Overall our QW realization has the major features expected of an ideal QW, with a momentum distribution which increases ballistically growing linearly with the number of steps $j$.

As a consequence of the relatively large range of walk steps in our scheme, we are also able to observe the quantum-to-classical transition by the addition of noise to each coin toss. This noise takes the form of a controllable randomization of the phase of each MW pulse. Figures~\ref{fig1}(c) to \ref{fig1}(f) demonstrate our experimental implementation of the quantum-to-classical transition for several different amounts of coin-phase noise; Fig.~\ref{fig1}(c) represents the standard QW of Fig.~\ref{fig1}(a) with coin phases fixed at $\chi=-\pi/2$. As mentioned, this QW is associated with the characteristic standard deviation of the momentum distribution ${\propto}j$. Signatures of a classical walk start to emerge by adding as little as $8\%$ randomness (with uniform distribution) to these phases, see Fig.~\ref{fig1}(d). Note how the ballistic peaks become less prominent after a few steps and that a Gaussian-like peak starts to emerge in the center. The Gaussian distribution of the walk with the characteristic standard deviation growing as ${\propto}\sqrt{j}$ is a manifestation of a classical walk. The walk becomes dominantly classical (Gaussian with no QW ballistic peaks) at $20\%$ phase randomness (e) and fully classical (f) when the phase is uniformly randomized within a full $2\pi$. The appearance of a Gaussian-like peak as a result of the noise enhancement is quite evident in Fig.~\ref{fig1}(g), which presents the momentum distributions at the eighth step of the walk.

\begin{figure}
	\begin{center}
		\centering
		\includegraphics[scale=0.055]{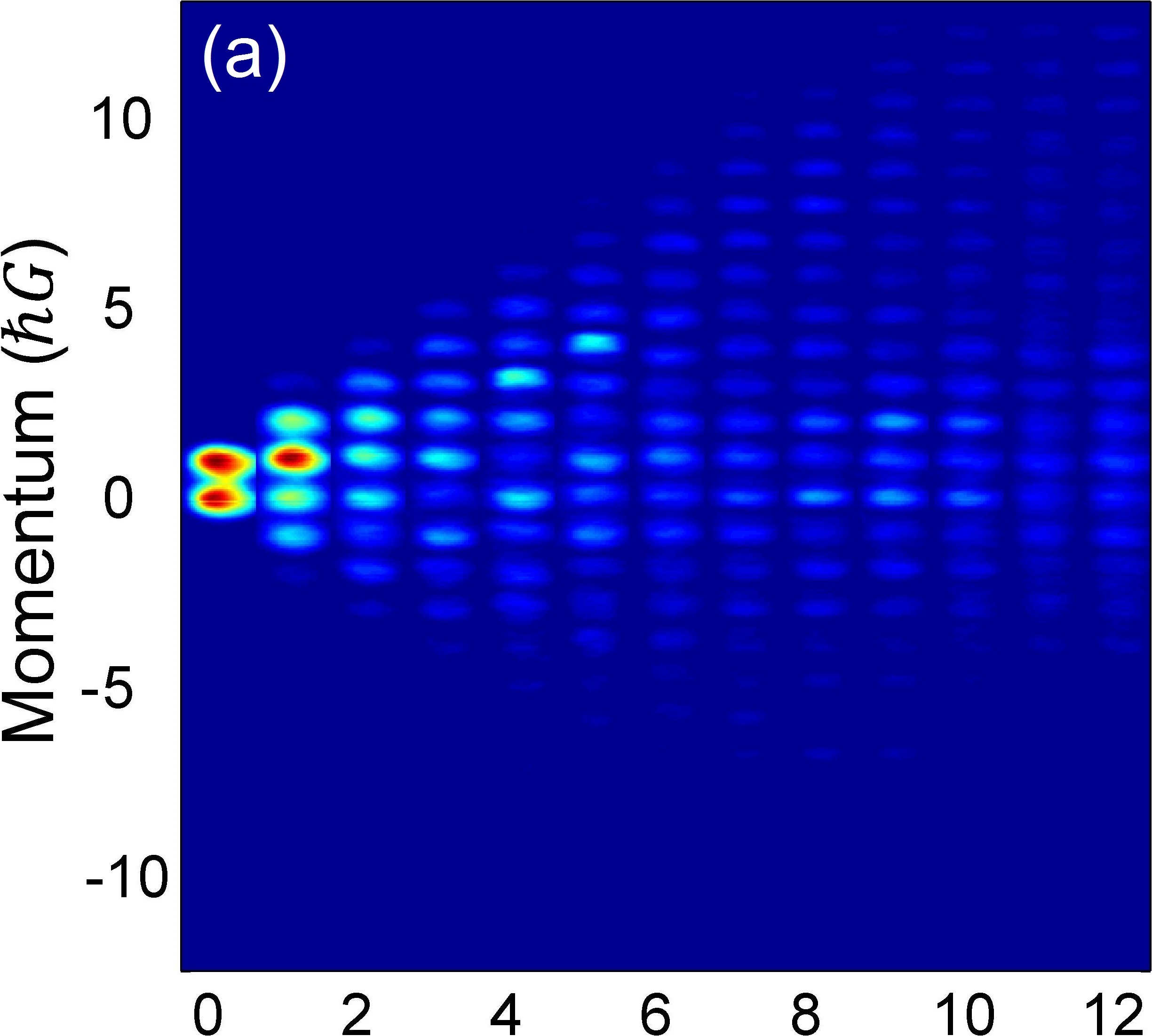}
		\vspace{0.2cm}
		\hspace{0.05cm}
		\includegraphics[scale= 0.055]{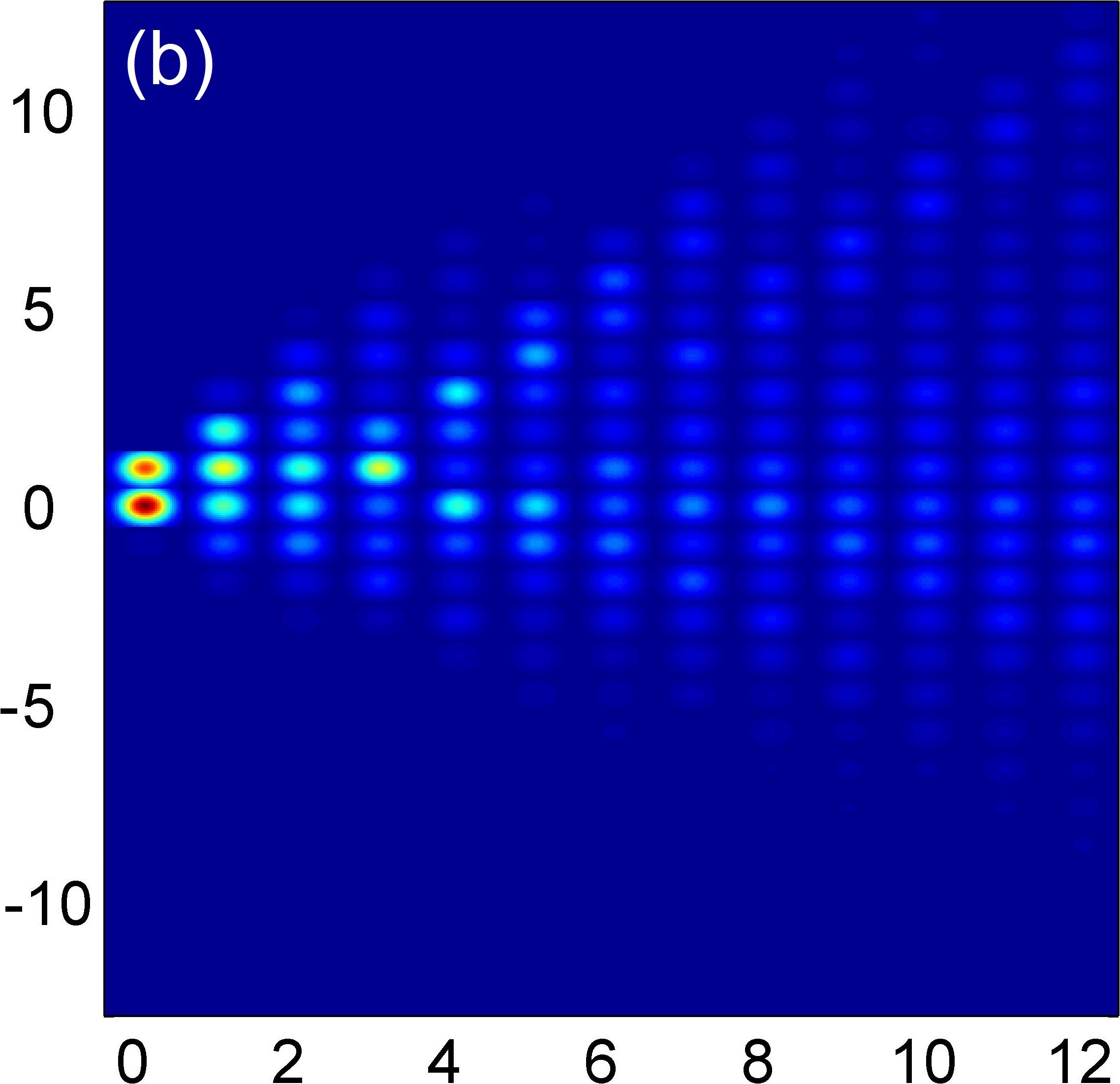}\\
		\includegraphics[scale=0.054]{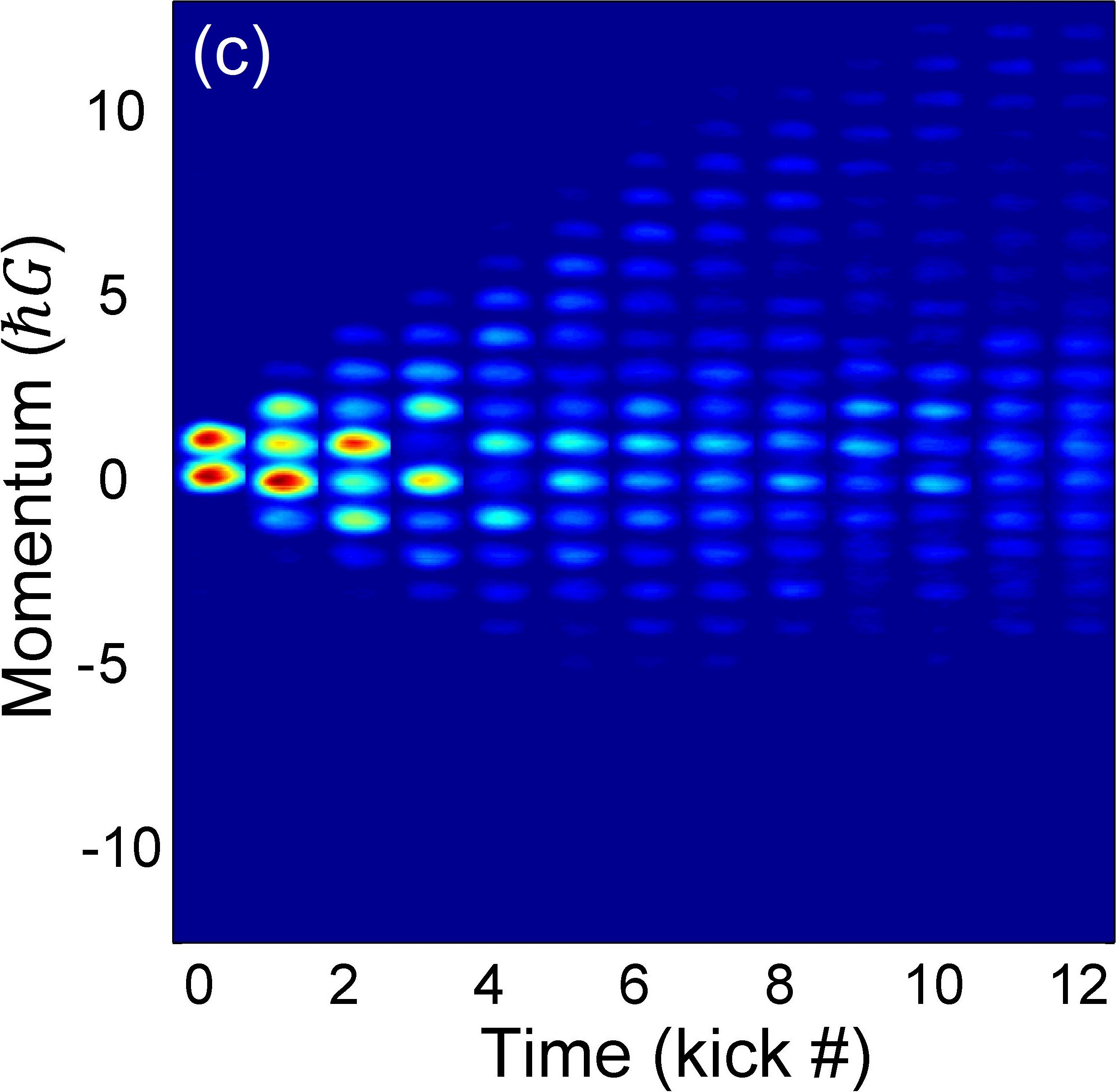}
		\hspace{0.05cm}
		\includegraphics[scale=0.0545]{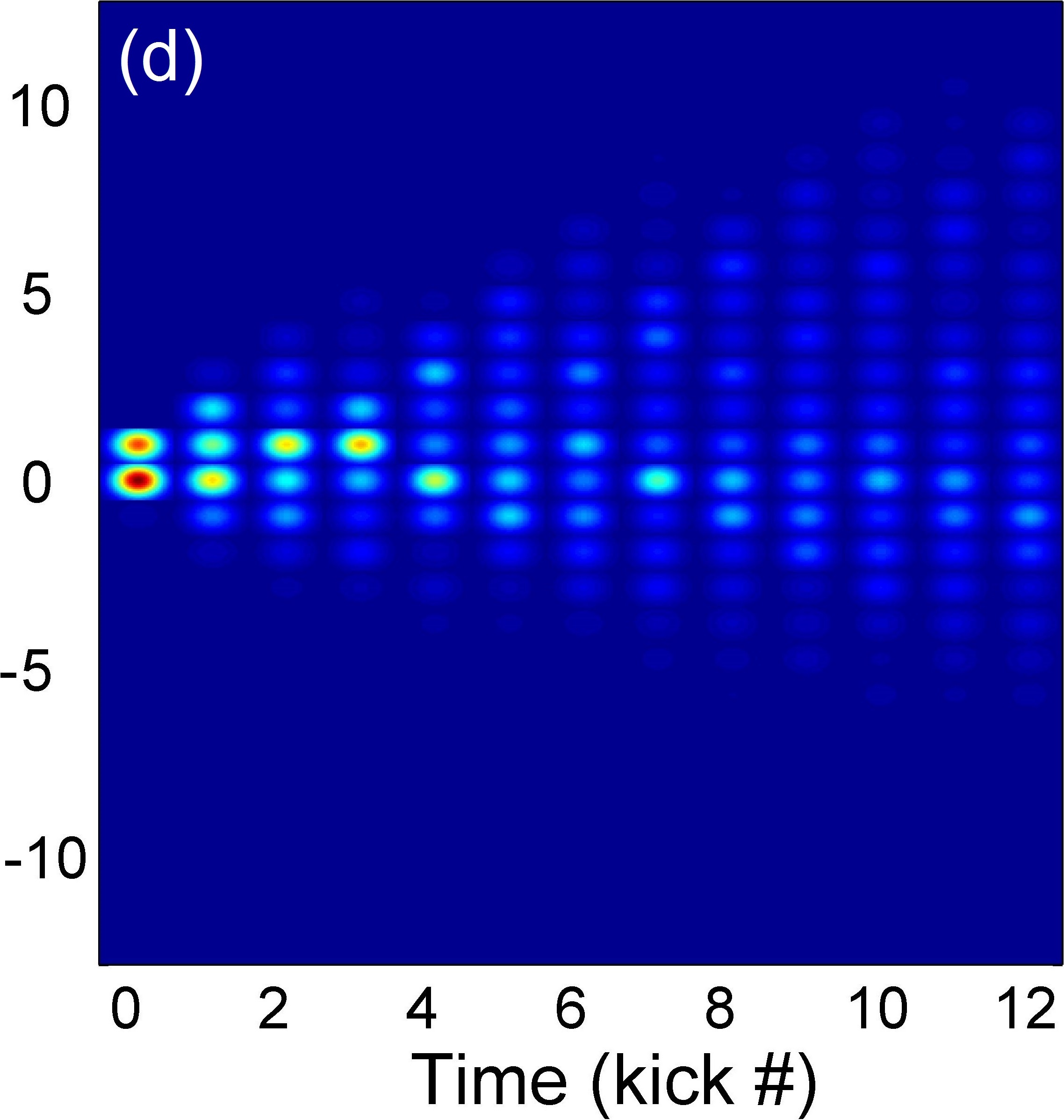}\\
		\vspace{0.2cm}
		\includegraphics[width=0.7\linewidth]{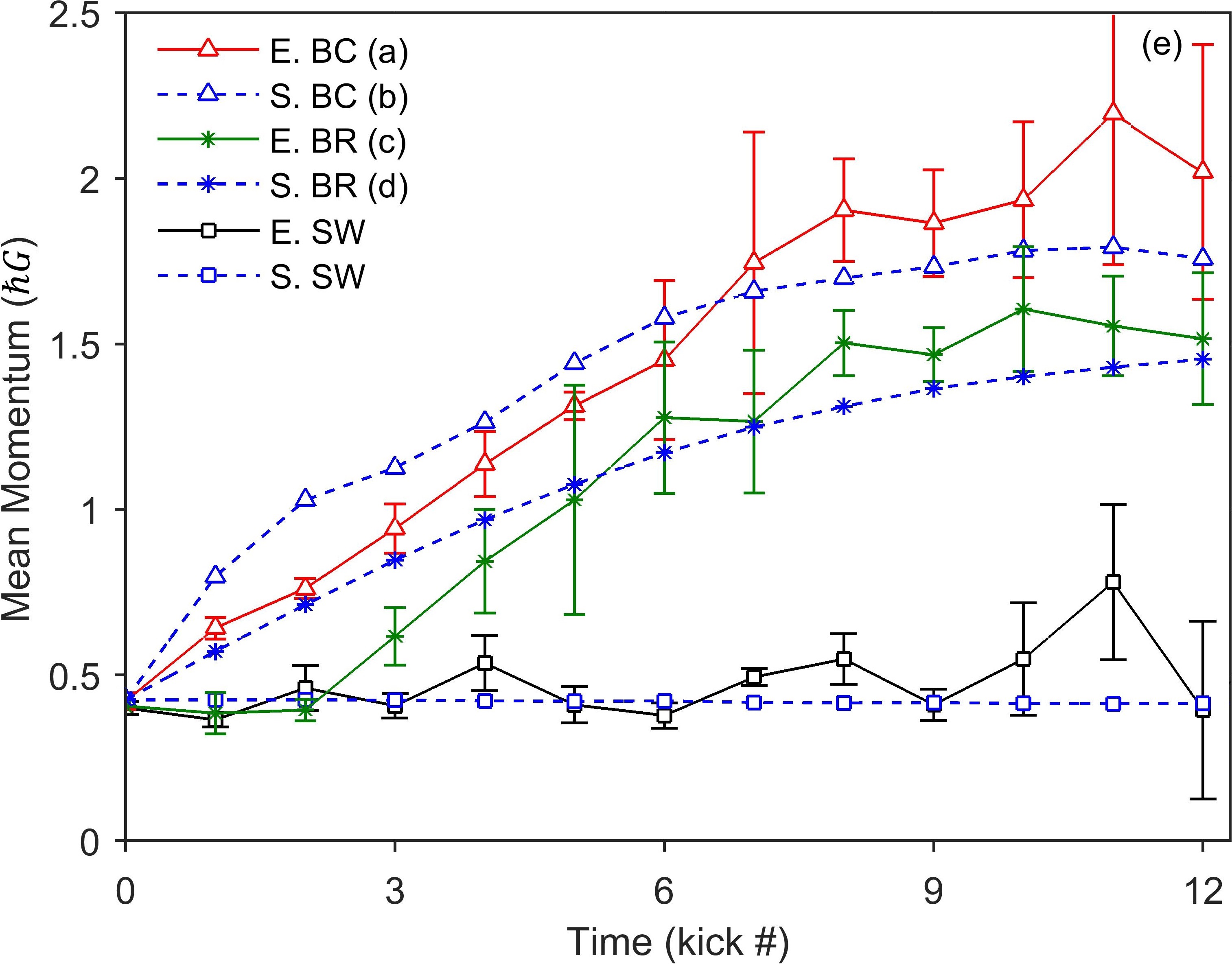}
		\caption{Experimental (a) and simulated (b) momentum distributions of the steered walk for a BC with $|k|=1.45$. Here the coin tosses were changed so as to produce the internal state ${\sqrt{0.7}|1{\rangle}+\sqrt{0.3}i|2{\rangle}}$ rather than ${\sqrt{0.5}(|1{\rangle}+i|2{\rangle}})$. Panels (c) and (d) demonstrate the experimental and simulated BR steered walks with an unbiased coin and $k_{1}=-1.7$, $k_{2}=+1.0$ instead of $k_{1}=-1.45$, $k_{2}=+1.45$. (e) shows the experimental (E.) and simulated (S.) variation of the mean momentum for BC and BR walks compared to the symmetric walk (SW).}	\label{fig2}
	\end{center}
\end{figure}

Our protocol also permits us to investigate biased QWs implemented through either a biased coin (BC) or via the use of nonsymmetric walk steps (i.e., the walk’s left and right shifts are not identical). We realize the former by altering the power of the MW pulses from the ${\pi}/2$ scheme so that unequal superpositions of internal states are obtained. In the latter case, a biased ratchet (BR) is achieved by detuning the ratchet laser so that the laser frequency is no longer halfway between the ground state hyperfine levels. Since the ratchet strength is inversely proportional to the detuning, this shift in the frequency results in unequal ratchet potentials for each state. Figure~\ref{fig2} demonstrates the experimental and simulated results of the steered walks for both BC and BR cases. As can be inferred, both the direction and speed of the walk can be manipulated by biasing the coin and ratchets in these ways. The controllability of the walk direction is particularly interesting for matter-wave interferometry and quantum search applications.

\begin{figure}
	\begin{center}
		\centering
		\includegraphics[scale=0.0552]{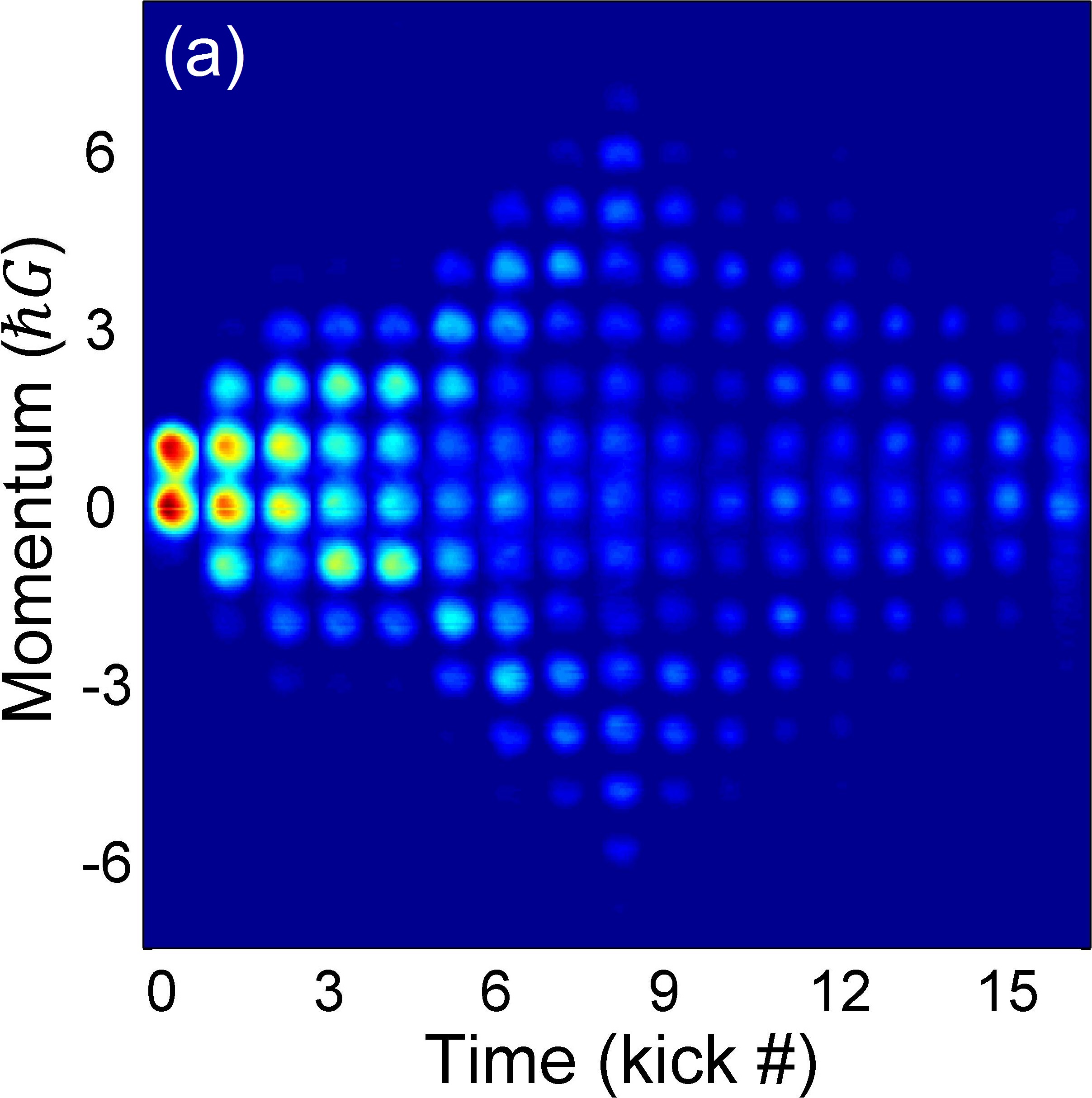}
		\hspace{0.1cm}
		\includegraphics[scale= 0.055]{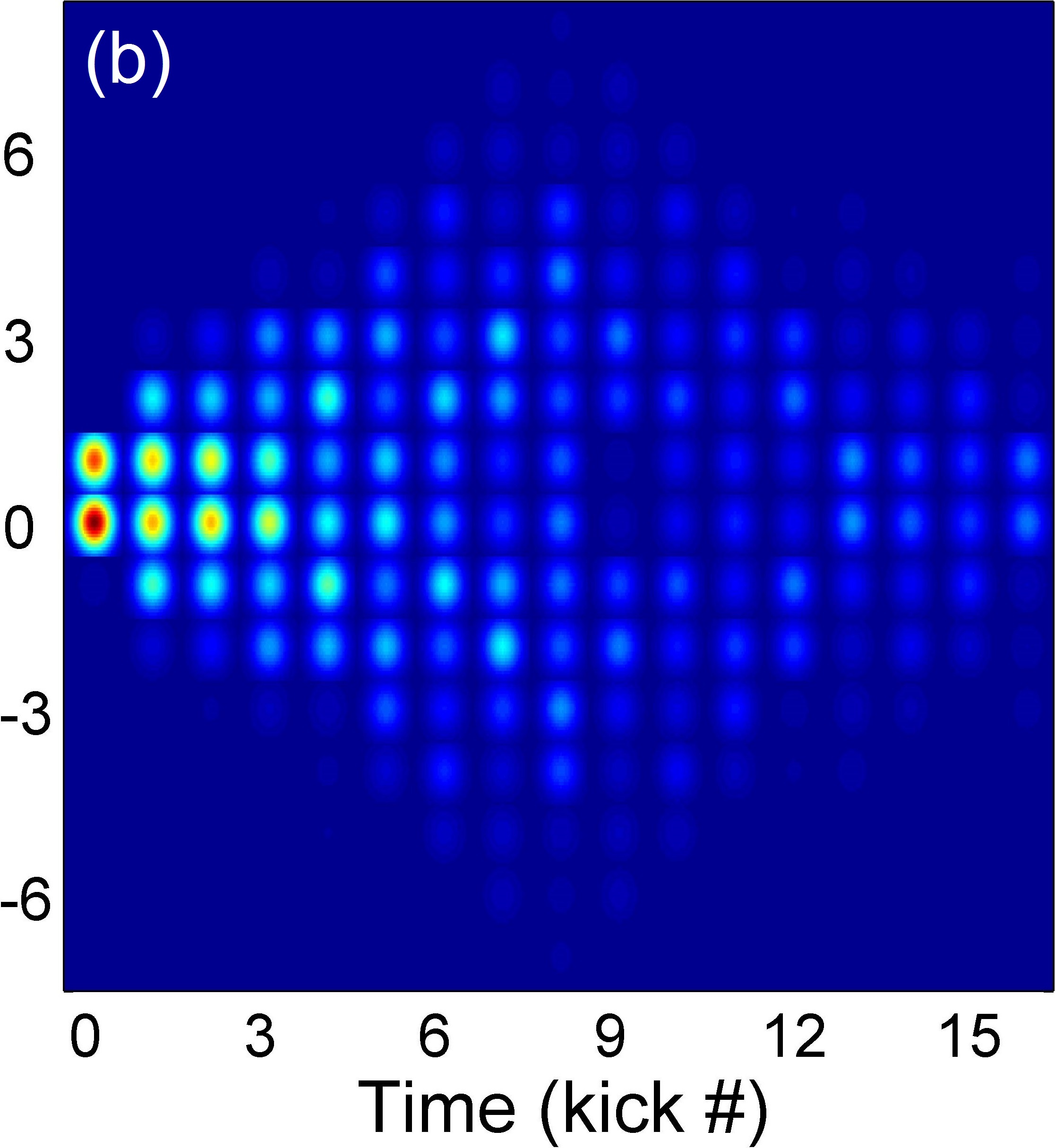}\\
		\vspace{0.2cm}
		\includegraphics[width=0.7\linewidth]{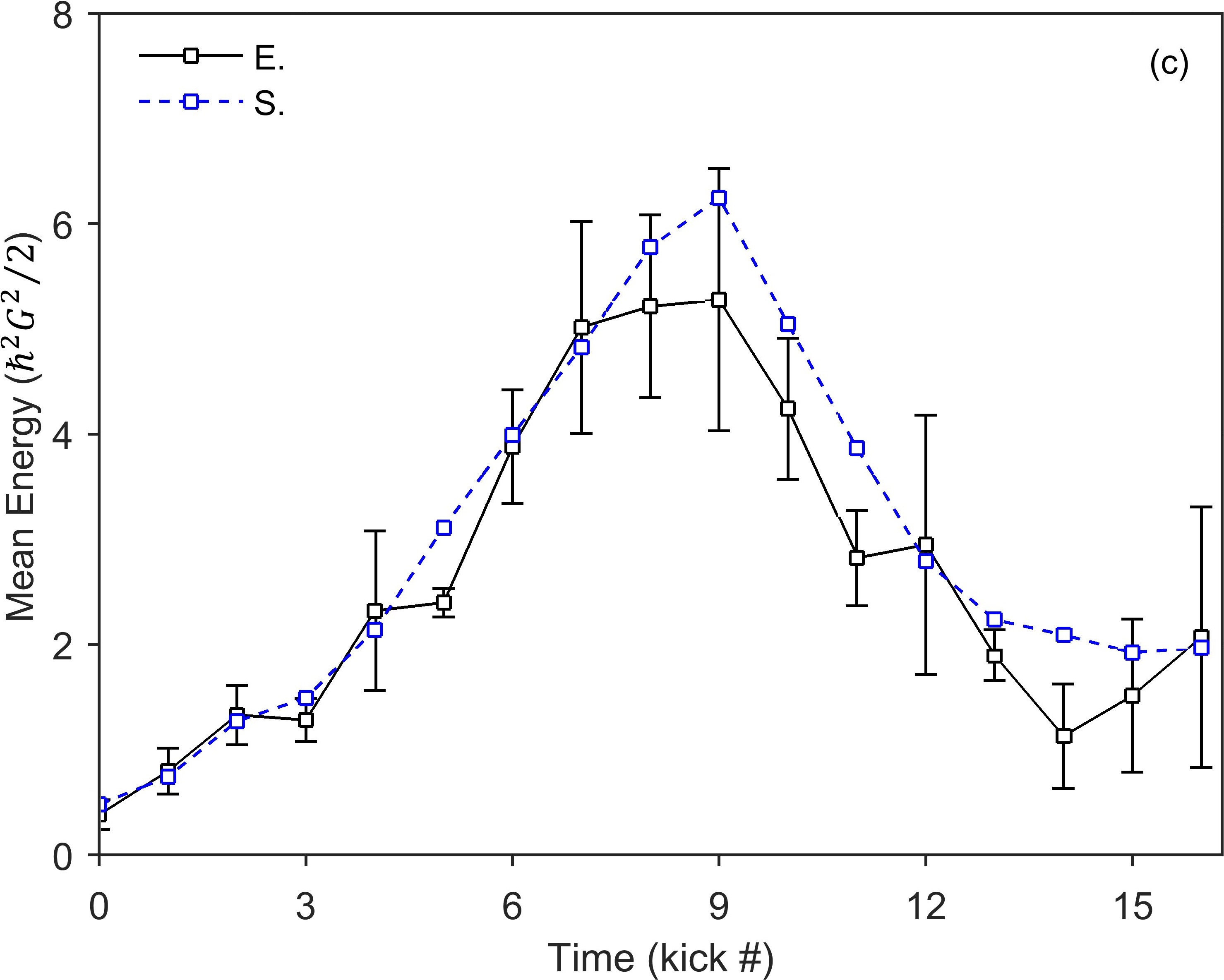}
		\caption{Experimental (a) and simulated (b) momentum distributions of a reversed QW with $|k|=1.45$. After eight of the walk steps ${\hat{\bf{U}}}_{\rm{step}}={\hat{\bf{T}}{\hat{\bf{M}}}}$ are taken, applying their Hermitian conjugate ${{\hat{\bf{U}}}}^\dagger_{\rm{step}}={{\hat{\bf{M}}}}^\dagger{{\hat{\bf{T}}}}^\dagger$ for an additional eight steps reverts the system to its original state. These conjugates can be realized as $\hat{\bf{M}}^\dagger=\hat{\bf{M}}(\pi/2,\pi/2)$ and $\hat{\bf{T}}^\dagger=\hat{\bf{M}}(\pi,\pi/2)\hat{\bf{T}}\hat{\bf{M}}(\pi,-\pi/2)$ that when the intermediate ${\hat{\bf{M}}}$ matrices multiply for successive steps, the reversal steps boil down to ${{\hat{\bf{U}}}}^\dagger_{\rm{\small step}}=\hat{\bf{T}}\hat{\bf{M}}(\pi/2,\pi/2)$ after a $\hat{\bf{T}}\hat{\bf{M}}(\pi,-\pi/2)$ ``reflection''. Panel (c) shows the experimental (E.) and simulated (S.) mean energy for a reversed walk with the minimum possible quasimomentum width and thermal cloud.}\label{fig3}
	\end{center}
\end{figure}

Two of the principal features of QWs that distinguish them from their classical counterparts are a unitary evolution and an entanglement between the internal and external d.o.f. These properties can be used to reverse a QW so as to retrieve the initial state of the system. We realize such a QW reversal by applying the Hermitian conjugates of the steps taken; i.e., ${{\hat{\bf{U}}}}^\dagger_{\rm{step}}={{\hat{\bf{M}}}}^\dagger{{\hat{\bf{T}}}}^\dagger$. The results of our QW reversal are shown in Fig.~\ref{fig3}. Our reversible QW could be of use in atom interferometry. The interference signal of the recombining momentum currents is determined by their phase difference and can be used as a sensitive measure of any perturbation affecting the phase of the system. The ideal dynamics of the walk and hence the fidelity of reversal is also sensitive to nonideal quasimomenta (not fulfilling the quantum resonance conditions necessary for the ratchet dynamics \cite{SW2011,Gil2013b,Ann2017}) and the amount of thermal cloud in the BEC.
 
We have reported on the realization of a fully controllable QW in momentum space with ultracold $^{87}$Rb atoms. With our present setup, we can experimentally implement QWs up to 20 steps, a range which is sufficiently large to observe the quantum-to-classical transition. This can be improved by relatively minor changes to our atom detection system so that a wider range of momenta can be observed, possibly up to the order of about 100 momentum classes before other experimental limitations become important. The latter include a breakdown of the Raman-Nath regime because of the finite pulse width \cite{rasel1995atom,klappauf1999quantum,oberthaler1999observation}, decoherence by spontaneous emission \cite{casper}, and vibrations of the optical setup. QWs with less diffusion between the ballistically spreading momentum currents could be achieved through the choice of an initial state composed of more momentum states \cite{Ann2017}.

We can control the directionality of our walks by manipulating the superposition of the internal states (BC) or by changing the relative detunings from the hyperfine levels (BR). This should allow for the compensation of probable biases in the dynamics of the quantum transport. Moreover, owing to the unitary nature of the QW and the entanglement between the internal and external d.o.f., we are able to reverse the walk. Using this feature, we can retrieve the quantum information encoded in previous steps of the walk or even recover the initial state of the system. This reversible QW may also be useful as an atomic interferometer since the reversibility is extremely sensitive to phases \cite{PNAS2012, Topo2016, Topo2, Kase2015N, Bragg2016}.

As a result of the fact that the QW takes place in momentum space \cite{Kase2015N, Ann2017,Bragg2016}, our walk should be readily extendable to higher dimensions \cite{Science2012, PhysRevLett.112.143604, sadgrove2012}. Multidimensional walks could be implemented using lattices with more than a single spatial dimension, or perhaps more straightforwardly, by the introduction of additional spatial frequency components to the one-dimensional lattice that is the basis of the momentum shift operator. Because of the BEC nature of our walker, unlike the single-particle systems \cite{meschede2009, PhysRevLett.103.090504, PhysRevLett.104.153602, PhysRevLett.104.100503}, our study can be extended to realize many-body walks by taking atom-atom interactions \cite{Preiss2015,Spin1,AW2017} into account. Further applications in driven walks \cite{PhysRevA.82.033602, NJP2016} and quantum algorithms, e.g., searches of marked momentum states, as in \cite{sadgrove2012} but by adjusting the coin degree, seem possible.

\textbf{Acknowledgments:}
We thank E. Arimondo, M. Sadgrove, A. Sundararaj and J. Clark for helpful comments and discussions.


\bibliography{ref.bib}

\end{document}